\documentclass[aps,prd,10pt,twocolumn,superscriptaddress,showpacs]{revtex4-1}
\usepackage{titlesec}
\usepackage{hyperref}
\usepackage{graphicx}
\usepackage{amsfonts,amsmath,amssymb,bm,bbm}
\usepackage{color}

\hypersetup{
    pdfnewwindow=true,      
    colorlinks=true,       
    linkcolor=blue,          
    citecolor=blue,        
    filecolor=blue,      
    urlcolor=blue        
}

\titleformat{\section}[runin]
  {\normalfont\bfseries}{\thesection}{1em}{}[:]
\titlespacing*{\section}{0cm}{2em}{1em}
\titleformat{\subsection}[runin]
  {\normalfont\itshape}{\thesubsection}{1em}{}[:]
\titlespacing*{\subsection}{1cm}{2em}{1em}

\begin{document}

\title{IC at IC: IceCube can constrain the intrinsic charm of the proton}
\author{Ranjan Laha}
\affiliation{Kavli Institute for Particle Astrophysics and Cosmology (KIPAC),\\ Department of Physics, Stanford University, Stanford, CA 94035, USA\\
SLAC National Accelerator Laboratory, Menlo Park, CA 94025, USA}

\author{Stanley J. Brodsky}
\affiliation{SLAC National Accelerator Laboratory, Stanford University, Stanford, CA 94025, USA \\
{\tt rlaha@stanford.edu,sjbth@slac.stanford.edu}}

\date{\today}

\begin{abstract}
The discovery of extraterrestrial neutrinos in the $\sim$ 30 TeV -- PeV energy range by IceCube provides new constraints on high energy astrophysics.  An important background to the signal are the prompt neutrinos which originate from the decay of charm hadrons produced by high energy cosmic-ray particles interacting in the Earth's atmosphere.  It is conventional to use the calculations of charm hadroproduction using gluon splitting $g \to c \bar c$ alone.  However, QCD predicts an additional ``intrinsic" component of the heavy  quark distribution which arises from diagrams where heavy quarks are multiply connected to the proton's valence quarks.  We estimate the prompt neutrino spectrum due to intrinsic charm.  We find that the atmospheric prompt neutrino flux from intrinsic charm is comparable to those calculated using QCD computations not including intrinsic charm, once we normalize the intrinsic charm differential cross sections to the ISR and the LEBC-MPS collaboration data.  In future, IceCube will constrain the intrinsic charm content of the proton and will contribute to one of the major questions in high energy physics phenomenology.
\end{abstract}


\maketitle

\section{Introduction}
\label{sec:Introduction}

Astrophysical neutrinos ($\equiv$ $\nu + \bar{\nu}$) discovered by IceCube provide new insights on profound astrophysics and particle physics questions\,\cite{Aartsen:2013bka,Aartsen:2013jdh,Aartsen:2014gkd,Aartsen:2015knd,Aartsen:2015rwa,Aartsen:2015zva,Aartsen:2016xlq}.  Many astrophysical models have been proposed  to explain these events\,\cite{Laha:2013eev,Murase:2013ffa,Anchordoqui:2013dnh,Kistler:2013my,Ahlers:2013xia,Chen:2013dza,Anchordoqui:2014pca,Fang:2014qva,Kashiyama:2014rza,Baerwald:2014zga,Chen:2014gxa,Ando:2015bva,Emig:2015dma,Tamborra:2015fzv,Murase:2015xka,Kistler:2015ywn,Kistler:2015oae,Kistler:2016ask,Moharana:2016mkl,Halzen:2016uaj,Pagliaroli:2016lgg,Anchordoqui:2016dcp,Dey:2016psn,Fang:2016hyv,Palladino:2016zoe,Senno:2015tsn} and to constrain various processes\,\cite{Bustamante:2015waa,Arguelles:2015dca,Vincent:2016nut,Kopp:2015bfa,Dasgupta:2012bd,Murase:2015gea,Rott:2014kfa,Esmaili:2014rma,Bhattacharya:2014vwa,Bhattacharya:2014yha,Ng:2014pca,Dutta:2015dka,Ko:2015nma,Aisati:2015ova,Boucenna:2015tra,Roland:2015yoa,Cherry:2016jol,Dev:2016uxj,Dev:2016qbd,Ema:2016zzu,DiBari:2016guw,Dey:2016sht,Vogel:2017fmc}. These have spurred development of new signatures such as the through-going tracks caused by $\tau$ leptons\,\cite{Kistler:2016ask} and the echo technique\,\cite{Li:2016kra}.

IceCube has detected an excess of neutrinos over the atmospheric neutrino background; however: how well do we know the background?  The contribution of conventional atmospheric neutrinos, produced from the decays of $\pi$'s and $K$'s, is known to $\sim$ 20\% -- 30\% precision depending on the energy\,\cite{Honda:2006qj,Barr:2004br,Fedynitch:2012fs}.   The major background uncertainty comes from $pp \rightarrow c X$, which results in prompt neutrinos produced from the decay of charm hadrons\,\cite{Bugaev:1989we,Gondolo:1995fq,Pasquali:1998ji,Gelmini:1999ve,Gelmini:1999xq,Candia:2003ay,Martin:2003us,Berghaus:2007hp,Enberg:2008te,Gaisser:2013ira,AhnICRC2013,Lipari:2013taa,Engel:2015dxa,Bhattacharya:2015jpa,Gauld:2015kvh,Gauld:2015yia,Garzelli:2015psa,Fedynitch:2015zma,Fedynitch:2015zbe,bausimpact,Bhattacharya:2016jce,Gaisser:2016obt,Halzen:2016pwl,Halzen:2016thi,Garzelli:2016xmx,Benzke:2017yjn}.  The flavor ratio of prompt neutrinos is $\nu_e : \nu_\mu : \nu_\tau \approx 1:1:0.1$, and $\nu : \bar{\nu} = 1 : 1$.

Most calculations of the prompt neutrino spectrum from charm hadroproduction are based within  perturbative QCD (pQCD) gluon splitting $g \to c\bar c$ alone\,\cite{Pasquali:1998ji,Gelmini:1999ve,Gelmini:1999xq,Candia:2003ay,Martin:2003us,Berghaus:2007hp,Enberg:2008te,Gaisser:2013ira,AhnICRC2013,Engel:2015dxa,Bhattacharya:2015jpa,Gauld:2015yia,Gauld:2015kvh,Garzelli:2015psa,Fedynitch:2015zma,Fedynitch:2015zbe,bausimpact,Bhattacharya:2016jce,Fedynitch:2016nup,Garzelli:2016xmx,Benzke:2017yjn}.  Inclusion of nonperturbative effects, for e.g., intrinsic charm\,\cite{Brodsky:1980pb,Brodsky:2015fna,Brodsky:2016fyh,Hobbs:2016hpe}, have received much less consideration\,\cite{Bugaev:1989we,Gondolo:1995fq}.  Recently attention has been drawn to forward production of charm hadrons by Refs.\,\cite{Halzen:2016pwl,Halzen:2016thi}.  Intrinsic charm is a rigorous prediction of QCD (see Supplementary Materials) and it is important to estimate its effect on atmospheric prompt neutrinos. 

The important distinction between intrinsic charm and gluon splitting is that intrinsic charm uses the incoming proton energy much more efficiently due to its harder $d\sigma/ dx_F$ distribution.  Inclusion of nonperturbative effects are important since the amount of intrinsic charm is an important uncertainty in QCD simulations.  Due to its inherent non-perturbative nature, it has not yet been calculated from first principles, and thus its normalization must be inferred from experiment.  Experiments have not yet decisively measured the normalization of intrinsic charm in the proton, which typically dominates the differential cross section at high $x_F$.

Various experimental techniques have been suggested for measuring atmospheric prompt neutrinos\,\cite{Gelmini:2002sw,Beacom:2004jb,Gandhi:2005at,Desiati:2010wt}.  These studies illustrate how measurements can constrain the underlying QCD mechanism in regions of the parameter space where it is difficult to obtain constraints from colliders\,\cite{bausimpact}.  

IceCube compares the prompt neutrino spectrum derived by Enberg, Reno and Sarcevic (with modifications by Gaisser) (ERS w/G)\,\cite{Enberg:2008te,Gaisser:2013ira,Gaisser:2016obt} with their data.  The present upper limits on the prompt neutrinos are near the nominal predictions\,\cite{Aartsen:2014muf,Aartsen:2016xlq,Aartsen:2015zva}.  An additional contribution to the prompt neutrino spectrum can change the interpretation of the astrophysical neutrinos. 

In this paper, we calculate the prompt neutrino contribution from intrinsic charm after normalizing the differential cross section to the ISR and the LEBC-MPS collaboration data\,\cite{Chauvat:1987kb,Ammar:1988ta}.  This contribution must be added to the $g\rightarrow c\bar{c}$ contribution to obtain the total atmospheric prompt neutrino spectrum.   We show that the prompt neutrino flux from intrinsic charm can be comparable to those calculated within QCD computations not including intrinsic charm.  The inclusion of this component as a background in the atmospheric neutrino flux can have important implications on the flux and spectral shape of the ``IceCube excess neutrinos".

We emphasize that IceCube can test these differential cross sections which have proven to be difficult to measure in colliders.   This synergy between IceCube and the collider searches\,\cite{Rostami:2015iva,Bailas:2015jlc,Boettcher:2015sqn,Lipatov:2016feu} can constrain the normalization of the intrinsic charm contribution and contribute to the investigation of a $\sim$36 year old puzzle in QCD.

\section{Calculations of neutrino fluxes}
\label{sec:calculation}

The earliest prompt neutrino calculations employed a proton-only cosmic ray flux known as the ``broken power-law"\,\cite{Gondolo:1995fq,Pasquali:1997cg,Pasquali:1998ji,Pasquali:1998xf,Enberg:2008te}.  Recent observations of cosmic ray flux indicate a mixed composition\,\cite{Gaisser:2011cc,Gaisser:2013bla,Stanev:2014mla}: the Gaisser 2012 fit\,\cite{Gaisser:2011cc} with $(i)$ the third component being proton (H3P), or $(ii)$ mixed (H3A), and the Stanev et al., 2014 fit\,\cite{Stanev:2014mla} fit with $(iii)$ three (H14A), or $(iv)$ four cosmic ray populations (H14B).  We  convert these to an equivalent all-proton flux, $\phi_p (E,X)$, where $E$ and $X$ denote the proton energy and the atmospheric column depth, respectively\,\cite{Gauld:2015kvh}.  

Assuming that the fluxes are separable in energy and column depth, we write the cascade equations as \,\cite{Gaisser:1990vg,Lipari:1993hd,Gondolo:1995fq,Pasquali:1998ji,Enberg:2008te,Gauld:2015kvh,Garzelli:2015psa}
\begin{eqnarray}
\dfrac{d\phi_p (E,X)}{dX} = -\dfrac{\phi_p (E,X)}{\lambda_p (E)} + Z_{pp} (E) \dfrac{\phi_p (E,X)}{\lambda_p (E)} \, ,
\label{eq:proton cascade equation}
\end{eqnarray}
\begin{eqnarray}
\dfrac{d\phi_m (E,X)}{dX} &=& -\dfrac{\phi_m (E,X)}{\rho(X) \, \delta_m(E)} -\dfrac{\phi_m (E,X)}{\lambda_m (E)} \nonumber\\
&+& Z_{mm} (E) \dfrac{\phi_m (E,X)}{\lambda_m (E)} + Z_{pm} (E) \dfrac{\phi_p (E,X)}{\lambda_p (E)} \,, \phantom{111}
\label{eq:meson cascade equation}
\end{eqnarray}
\begin{eqnarray}
\dfrac{d\phi_\ell (E,X)}{dX} = \sum_m Z_{m\ell}(E) \dfrac{\phi_m (E,X)}{\rho(X) \, \delta_m(E)} \, ,
\label{eq:lepton cascade equation}
\end{eqnarray}
where $\lambda_p (E)$ [$\lambda_m (E)$] denotes the nucleon [charm hadron] attenuation length.  The charm hadron flux [lepton flux from the decay of charm hadron] are denoted by $\phi_m (E,X)$ [$\phi_\ell (E,X)$].  The atmospheric density and charm hadron decay length is denoted by $\rho(X)$ and $\delta_m (E)$, respectively.  The sum includes the contribution of all the relevant charm hadrons.

The production moments $Z_{pp} (E)$, $Z_{mm} (E)$, and $Z_{pm} (E)$ are defined as\,\cite{Pasquali:1998ji}
\begin{eqnarray}
Z_{kj}(E) = \int _0 ^1 \dfrac{dx_E}{x_E} \dfrac{\phi_k \left(\dfrac{E}{x_E},0 \right)}{\phi_k (E,0)} \dfrac{\lambda_k(E)} {\lambda_k \left(\dfrac{E}{x_E} \right)} \dfrac{dn_{kj}}{dx_E} (E/x_E)\,, \phantom{111}
\label{eq:production moment}
\end{eqnarray} 
where $x_E = E/E_k$, and $dn_{kj} (E/x_E)/ dx_E$ denote the production spectrum of $j$ from the interaction of $k$ with the air nucleon.  The decay moments $Z_{m \ell} (E)$ are calculated following Refs.\,\cite{Pasquali:1998ji,Enberg:2008te}.


\begin{figure}
\includegraphics[angle=0.0,width=0.5\textwidth]{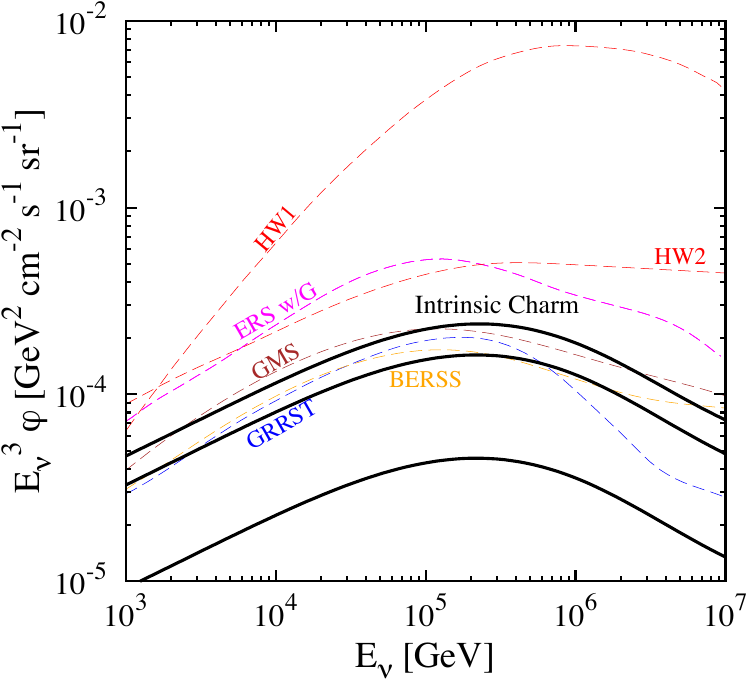}
\caption{Predictions for the atmospheric prompt neutrino ($\nu_e$ + $\bar{\nu}_e$ or $\nu_\mu$ + $\bar{\nu}_\mu$) spectrum, $\varphi$, as a function of the neutrino energy $E_{\nu}$ using the H3A cosmic ray input flux.  We show $(i)$ GRRST\,\cite{Gauld:2015kvh}, $(ii)$ BERSS\,\cite{Bhattacharya:2015jpa}, $(iii)$ GMS\,\cite{Garzelli:2015psa}, $(iv)$ ERS w/G\,\cite{Enberg:2008te,Gaisser:2013ira}, $(v)$ HW1\,\cite{Halzen:2016pwl}, $(vi)$ HW2\,\cite{Halzen:2016thi}, and $(vii)$ our calculation (Intrinsic Charm).  The highest, intermediate, and the lowest flux from the intrinsic charm contribution correspond to Case (A), Case (B), and Case (C) respectively.  See text for details.  The upper limit from the IceCube data on the prompt neutrino flux is 1.06 times the ERS w/G flux\,\cite{Aartsen:2016xlq}.}   
\label{fig:Comparison prompt spectrum}
\end{figure}


For $\lambda_p (E)$, we take the mean atomic number of air molecules, $\langle A \rangle = 14.5$.  For the proton - air cross section, we take the values from QGSJet0.1c\,\cite{Kalmykov:1993qe}.  Additional parameters required to calculate  $Z_{pp} (E)$, $Z_{mm} (E)$ and $\lambda_m (E)$ are taken from Refs.\,\cite{Bhattacharya:2015jpa, Pasquali:1998ji}.


\begin{figure*}[!thpb]
\centering
\includegraphics[angle=0.0,width=0.49\textwidth]{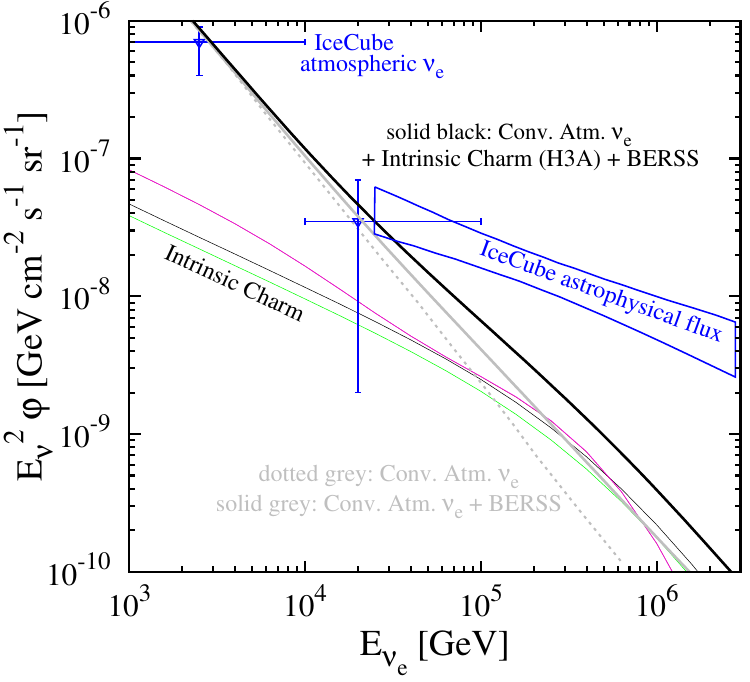}
\includegraphics[angle=0.0,width=0.49\textwidth]{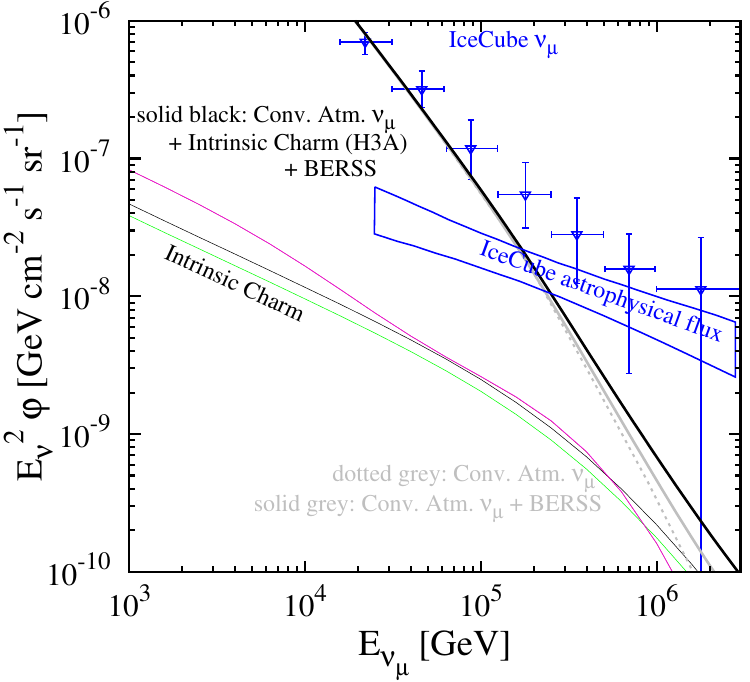}
\caption{{\bf Left:} Comparison of the total atmospheric $\nu_e$ + $\bar{\nu}_e$ data (IceCube-86 for 332 days) with calculations.  The contribution to the $\nu_e$ + $\bar{\nu}_e$ flux from intrinsic charm for Case (A) for various cosmic ray spectra is shown by the dashed lines (H3A = magenta, H3P = green, H14A = brown, and H14B = magenta.  H14A and H14B are on top of each other).  The conventional $\nu_e$ + $\bar{\nu}_e$ flux\,\cite{Aartsen:2015xup}, conventional $\nu_e$ + $\bar{\nu}_e$ + BERSS (H3A), and conventional $\nu_e$ + $\bar{\nu}_e$ + BERSS + intrinsic charm contribution for H3A are shown.  {\bf Right:} Same as the left panel, but for $\nu_\mu$ + $\bar{\nu}_\mu$\,\cite{Aartsen:2015zva} (IceCube-79/ 86 for 2 years).  This measurement also includes the astrophysical neutrino flux.  The astrophysical flux shown in these panels is from Refs.\,\cite{Aartsen:2015knd}.}
\label{fig:Atmospheric neutrino measurement}
\end{figure*}


The calculation of $Z_{pm} (E)$ involves the differential cross section $\dfrac{d\sigma}{dx_F}(pp \rightarrow cX)$.  There are substantial uncertainties in this differential cross section, especially at high $x_F$.  Modern colliders are not capable of measuring this differential cross section in the forward region (high $x_F$)\,\cite{bausimpact}.  State of the art calculations, which incorporate various different constraints, are also lacking for these differential cross sections at high $x_F$.  Taking these uncertainties into account, we adopt three test cases using the data presented by the ISR experiments and the LEBC-MPS collaboration. 

Case (A):  For $\Lambda_c$ production, we use Ref.\,\cite{Hobbs:2013bia} which normalizes their differential cross section to the ISR data\,\cite{Chauvat:1987kb}.  For $D$ mesons, we use the shape of the differential cross sections as calculated in Ref.\,\cite{Gutierrez:1998bc}, and normalize them to the data at the highest $x_F$ ($d\sigma/dx_F$ $\approx$ 17$^{+18}_{-9} \, \mu$b at $x_F$ $\approx$ 0.32) as measured by the LEBC-MPS collaboration\,\cite{Ammar:1988ta}.    

Although the LEBC-MPS measurements extend to the forward region, $x_F \approx 0.32 \pm 0.08$, yet due to the uncertainties in the theoretical prediction and experimental measurements, it is difficult to estimate the contribution of intrinsic charm from this.  Our strategy is to use the best-fit prediction following Ref.\,\cite{Bhattacharya:2015jpa}  ($\approx$ 10 $\mu$b  at $x_F$ = 0.32), and use the error bars of the LEBC-MPS measurement to maximize the intrinsic charm contribution.  This sets the normalization for the $D$ mesons.

Case (B):  We use the charm hadron differential cross section spectral shapes as derived by Ref.\,\cite{Gutierrez:1998bc}.  To normalize these, we assume that the intrinsic charm cross section $d\sigma/dx_F$  $\approx$ 25 $\mu$b at $x_F \approx 0.32$ for the $D$ mesons.  Since we are using the same model for the $D$ mesons and the $\Lambda_c$ production, this also gives the normalization of the $\Lambda_c$ cross section.

Case (C):  We again use the charm hadron differential cross section spectral shape as derived by Ref.\,\cite{Gutierrez:1998bc}.  We normalize the cross section such that the intrinsic charm cross section $d\sigma/dx_F \approx$ 7 $\mu$b at $x_F \approx 0.32$ for the $D$ mesons.  This corresponds to the best fit point of the LEBC-MPS measurement.  Similar to Case (B), we use the same production model for $D$ mesons and $\Lambda_c$ production.  The above mentioned differential cross section for the $D$ mesons also set the normalization for the $\Lambda_c$ production.

The cross sections for various charm mesons and hadrons at $\sqrt{s} \approx 39$ GeV for the three cases discussed above are given in Table\,\ref{tab:cross sections}.  The cross section for the production of $\Lambda_c^+ + \Lambda_c^-$ in Case A is anomalously high as the normalization is fit to the ISR data.  The production cross section for these charmed bound states in the other cases are $\sim \mathcal{O}(\mu$b) and these are normalized to the LEBC-MPS data.    

We illustrate the uncertainty of the intrinsic charm flux by the three cases as mentioned above.  The Case (C) does not represent a lower limit to the intrinsic charm contribution to the differential cross section.  It is possible that the intrinsic charm contribution is lower, and this will correspond to a lower contribution to the atmospheric prompt neutrino flux compared to what is presented here.  As is evident from our discussion, the intrinsic charm cross section is not at all well known.  Despite decades of effort, colliders have not yet been able to definitively measure its normalization.  We examine the role of IceCube in this search.  

The intrinsic charm cross section scales with the mass number, $A$, approximately as $A^{0.755 \pm 0.016}$, according to SELEX\,\cite{BlancoCovarrubias:2009py}.  The energy dependence of the intrinsic charm contribution follows the inelastic cross section\,\cite{Antchev:2011vs}.

We solve Eqs.\,\ref{eq:proton cascade equation} -- \ref{eq:lepton cascade equation} separately in the low and high energy regime\,\cite{Gondolo:1995fq,Pasquali:1998ji,Enberg:2008te,Gauld:2015kvh,Garzelli:2015psa}.  The final prompt neutrino flux is a geometric interpolation of the low and high energy solutions and includes the contribution of all the charm hadrons, $D^0, \bar{D}^0, D^\pm, D_s^\pm, \Lambda_c^\pm$.  The $\Lambda_c$ shares a c-quark from the $|$uudc$\bar{\rm c}$$\rangle$ state of the proton, and thus has a harder differential cross section $d\sigma/d x_F$ when compared to that of the $D$ mesons.

\begin{table}[b]

\caption{Production cross sections (in $\mu$b) at $\sqrt{s} \approx 39 \,{\rm GeV}$ for the various charm mesons and baryons via the intrinsic charm models discussed in the text.  Here $\sigma (D_{(s)}^\pm) = \sigma (D_{(s)}^+) + \sigma(D_{(s)}^-)$ and $\sigma (\Lambda_c^\pm) = \sigma (\Lambda_c^+) + \sigma (\Lambda_c^-)$.}
\begin{ruledtabular}
%
\begin{tabular}{lcccc}
Case & $ \sigma (D^\pm)$\, &  $\sigma (D^0 + \bar{D}^0)$\,  &\,  $ \sigma (D_s^\pm)$ \,&\, $\sigma (\Lambda_c^\pm)$ \\ 
\hline
Case A & 3.88 & 3.88 & 2.29 & 183.54\\
Case B & 3.88 & 3.88 & 2.29 & 4.99 \\
Case C & 1.09 & 1.09 & 0.64 & 1.47 \\
\end{tabular}
%
\end{ruledtabular}
\label{tab:cross sections}
\end{table}

Our calculation improves over the previous estimates\,\cite{Bugaev:1989we,Gondolo:1995fq,Fedynitch:2016nup,Halzen:2016pwl,Halzen:2016thi} in various important ways.  We normalize our calculations to the ISR and the LEBC-MPS collaboration data\,\cite{Chauvat:1987kb,Ammar:1988ta}, which were not used in the earliest works.  We employ the latest cosmic ray flux measurement, and the experimentally measured nuclear scaling of the cross section, and a theoretically motivated energy dependence of the cross section.  We use a more updated calculation of the intrinsic charm cross section which takes into account the inherent non-perturbativeness of the process\,\cite{Hobbs:2013bia,Gutierrez:1998bc}, whereas some of these earlier works\,\cite{Halzen:2016pwl,Halzen:2016thi} used a modified pQCD prescription to account for the high $x_F$ data.

\section{Results}
\label{sec:results}

Our predictions for the flux of neutrinos ($\nu_\mu + \bar{\nu}_\mu$ or $\nu_e + \bar{\nu}_e$) are shown in Fig.\,\ref{fig:Comparison prompt spectrum}.  The three flux scenarios are given by Case (A), Case (B), and Case (C).  We also show the best-fit flux calculated by BERSS\,\cite{Bhattacharya:2015jpa}, GMS\,\cite{Garzelli:2015psa}, GRRST\,\cite{Gauld:2015kvh}, HW1\,\cite{Halzen:2016pwl}, HW2\,\cite{Halzen:2016thi}, and ERS w/G\,\cite{Aartsen:2014muf,Aartsen:2015zva}, all of which have large theoretical error bars.  Remarkably, we find that the atmospheric prompt neutrino flux due to intrinsic charm can be at the same level as those estimated within QCD calculations not including intrinsic charm.

The neutrino fluxes due to intrinsic charm can be large enough to be detectable by IceCube.  The detectability depends on the contribution of the fluxes arising from QCD not including intrinsic charm.  For example, if the best fit prediction follows the BERSS flux, and the intrinsic charm contribution is as large as Case (B), then IceCube can get a strong constraint on intrinsic charm provided the uncertainties in BERSS are made smaller.  It is important to decrease the uncertainties within the various pQCD predictions in order to obtain a robust constraint on intrinsic charm; global analyses of laboratory data also have this same requirement\,\cite{Dulat:2013hea,Ball:2016neh}. 

For case (A), the dominant contribution to the flux comes from the production of $\Lambda_c^\pm$, followed by $D^\pm$, $D^0 + \bar{D}^0$, and $D_s^\pm$.  For cases (B) and (C), the production of $D^\pm$, $D^0 + \bar{D}^0$, $D_s^\pm$, and $\Lambda_c^\pm$ contribute to the atmospheric prompt neutrino flux in decreasing order.  These contributions can be simply understood by comparing their respective production cross sections (see Table\,\ref{tab:cross sections}) multiplied by the decay branching fractions to neutrinos.

If the intrinsic charm contribution follows Case (C), and the non intrinsic charm contribution follows the BERSS flux, then it will be difficult to measure the intrinsic charm unless very precise measurements of the atmospheric prompt neutrino fluxes are made.  Even in this pessimistic case, weak upper limits on intrinsic charm can be obtained from the data.  Encouragingly, present IceCube constraints have already started to constrain the forward production within the various QCD computations not including intrinsic charm\,\cite{Bhattacharya:2016jce,Garzelli:2016xmx,Benzke:2017yjn}.  It is expected that near future measurements of the prompt atmospheric neutrino flux will further constrain these various contributions in pQCD.

In the intrinsic charm picture, the proton preferentially forms a charm hadron with a similar energy.  In the $g\rightarrow c \bar{c}$ picture, due to its steeply falling $d\sigma/dx$ distribution, the charm hadron comes dominantly from a proton at much higher energy.  Our results are slightly lower than the calculation presented in Ref.\,\cite{Halzen:2016thi} due to the above mentioned refinements.

So far, IceCube has presented upper bounds on prompt neutrinos.  IceCube assumes that the prompt neutrino flux is the ERS w/G spectrum and varies the normalization.  The present limit on the prompt neutrino spectrum is 1.06 times the ERS w/G flux\,\cite{Aartsen:2016xlq}.  These IceCube limits are close to the intrinsic charm prompt neutrino spectrum predictions, implying that IceCube can give information about intrinsic charm content of the proton in the near future.

In Fig.\,\ref{fig:Atmospheric neutrino measurement} (left), we compare our calculation for Case (A) and the measurement of the atmospheric $\nu_e$ flux\,\cite{Aartsen:2015xup}.  The conventional atmospheric $\nu_e + \bar{\nu}_e$ flux (angular averaged) is taken from Refs.\,\cite{Honda:2006qj,Barr:2004br,Aartsen:2015xup}.  The conventional atmospheric $\nu_e$ + BERSS flux, the prompt $\nu_e$ flux due to intrinsic charm in case (A) for various different input cosmic ray model, the total atmospheric $\nu_e$ flux including the BERSS and due to intrinsic charm in case (A) for the H3A cosmic ray input model are also shown.  We also show the astrophysical neutrino spectrum from Ref.\,\cite{Aartsen:2015knd} in the energy range [25 TeV, 2.8 PeV].  This shows that although the inclusion of the intrinsic charm component can change the background for astrophysical neutrinos, yet atmospheric prompt neutrinos cannot explain the ``IceCube excess neutrinos".

Normalizing to the ISR and the LEBC-MPS collaboration data does not contradict the atmospheric $\nu_e$ measurements.  The importance of atmospheric $\nu_e$ measurement for prompt neutrinos was pointed out in Ref.\,\cite{Beacom:2004jb}, and we argue that it might be the best channel to search for intrinsic charm as well.  We predict that the atmospheric $\nu_e + \bar{\nu}_e$ flux due to intrinsic charm is larger than $g\rightarrow c \bar{c}$ contribution at $\gtrsim$ 50 TeV.  A more precise measurement of the atmospheric $\nu_e$ spectrum at slightly higher energies can give strong constraints on the intrinsic charm content of the proton.

Atmospheric prompt neutrinos cannot explain the ``IceCube excess neutrinos" since prompt neutrinos have a softer spectral shape and have accompanying muons.  The ``IceCube excess neutrinos" have an energy spectrum varying within $\sim$$E$$^{-2.1}$ and $E$$^{-2.6}$ between $\sim$30 TeV and 3 PeV and do not have any accompanying muons.  The prompt neutrino flux  follows the much softer cosmic ray spectrum.  

For downgoing events, the IceCube self veto can discriminate between atmospheric and astrophysical neutrinos\,\cite{Schonert:2008is,Gaisser:2014bja}.  Every atmospheric neutrino is accompanied by a muon or an electromagnetic shower from the same interaction producing the neutrino.  The muon or the shower detected in coincidence with the neutrino, reduces the atmospheric neutrino flux by a factor $\gtrsim$ 2 at energies $\gtrsim$ 10 TeV\,\cite{Aartsen:2014gkd}.  This also results in a difference in the zenith angle distributions of astrophysical and prompt neutrinos.

The angular distribution of atmospheric prompt neutrinos is approximately isotropic at $\lesssim$ 10$^7$ GeV.  Conventional atmospheric neutrinos have a smaller vertical flux compared to the  horizontal flux.  Searching for atmospheric neutrinos in the vertical direction can more easily find the prompt component.  More theoretical and experimental work is also required to narrow down the uncertainties of the predictions made within pQCD calculation to extract the contribution of intrinsic charm from the IceCube data.

A comparison of the $\nu_\mu + \bar{\nu}_\mu$ flux from the Northern Hemisphere with calculations is shown Fig.\,\ref{fig:Atmospheric neutrino measurement} (right)\,\cite{Aartsen:2015zva}.  The intrinsic charm component is shown for Case (A).  The astrophysical neutrino spectrum in the energy range [25 TeV, 2.8 PeV] from Ref.\,\cite{Aartsen:2015knd} is shown.  The neutrino flux due to intrinsic charm cannot increase, since it will be in contradiction with the ISR and the LEBC-MPS collaboration data.  The inclusion of this contribution may result in a revision of the astrophysical neutrino spectrum.  Since these events are up-going, the atmospheric veto does not play any role, and one needs to model the astrophysical neutrino flux before inferring the prompt neutrino contribution using this detection channel.

In Fig.\,\ref{fig:Atmospheric neutrino measurement}, we only show the intrinsic charm contribution to the prompt atmospheric neutrino flux for Case (A).  This has the largest flux among the three cases that we have considered, and hence we are displaying the optimistic case.  For this case, the total prompt atmospheric neutrino flux is dominated by the intrinsic charm contribution.  The intrinsic charm contribution in Case (B) is comparable to the best fit BERSS flux.  If the intrinsic charm contribution follows Case (C), then the total prompt atmospheric neutrino flux will be totally dominated by the BERSS flux.  In such case, upper limits on intrinsic charm of the proton can only be obtained if the prompt atmospheric neutrinos are measured quite precisely. 

We only plot the best fit BERSS flux in Fig.\,\ref{fig:Atmospheric neutrino measurement} for clarity.  We do not show the uncertainty in this flux which is substantial\,\cite{Bhattacharya:2015jpa}.  It is essential to decrease the uncertainties in this calculation to obtain a more robust constraint on intrinsic charm from astroparticle measurements.  

Present upper limits from IceCube have already started to constrain various pQCD computations not including intrinsic charm, and near future data will have stronger constraints\,\cite{Bhattacharya:2016jce,Garzelli:2016xmx,Benzke:2017yjn}.  There are spectral differences between astrophysical neutrinos and prompt atmospheric neutrinos.  IceCube uses a veto which produces a different angular dependence for the astrophysical neutrinos when compared to the prompt atmospheric neutrinos.  These distinct features help in determining the atmospheric prompt neutrino sample in the IceCube data.  Various different analyses of IceCube give similar upper limits on prompt atmospheric neutrinos implying that the constraint is robust.

\section{Conclusions}
\label{sec:conclusions}

The landmark discovery of astrophysical neutrinos by IceCube opens up a new era.  Due to the atmospheric veto employed by IceCube, any atmospheric neutrino spectrum shows an up v/s down asymmetry.  The excess of neutrinos unveiled by IceCube is isotropic implying the astrophysical origin of these events.  Careful consideration of the atmospheric neutrino background will impact the astrophysical neutrino flux interpretation.

The neutrino backgrounds considered so far by IceCube are the conventional atmospheric and prompt neutrinos predicted by $g\rightarrow c \bar{c}$.  Intrinsic charm, rigorously predicted by QCD, has strong theoretical justification and some experimental indications.  We find that this often neglected component can be as large as the component estimated within the various pQCD computations not including intrinsic charm, without violating any direct experimental constraints.  This has important implications in interpreting the astrophysical neutrino flux, and inferring the atmospheric prompt neutrino component.

We present our calculation of the neutrino flux due to intrinsic charm in Fig.\,\ref{fig:Comparison prompt spectrum} after normalizing to the ISR and the LEBC-MPS collaboration data.  We show the atmospheric prompt neutrino flux due to three different scenarios.  The atmospheric prompt neutrino flux due to intrinsic charm is comparable to that estimated within various pQCD computations not including intrinsic charm.  Our calculation is lower than Refs.\,\cite{Halzen:2016pwl,Halzen:2016thi} as we use improved theoretical and experimental input. 

The measurement of atmospheric $\nu_e + \bar{\nu}_e$ at higher energies is the most promising channel to discover prompt neutrinos and constrain the intrinsic charm of the proton (Fig.\,\ref{fig:Atmospheric neutrino measurement} left).  The comparison of the total atmospheric flux with the $\nu_\mu + \bar{\nu}_\mu$ data, including the intrinsic charm contribution, is shown in Fig.\,\ref{fig:Atmospheric neutrino measurement} (right).  The total atmospheric neutrino flux including intrinsic charm can dominate the flux contribution within the pQCD framework at energies $\gtrsim$ 200 TeV and $\gtrsim$ 2 PeV for $\nu_e + \bar{\nu}_e$ and $\nu_\mu + \bar{\nu}_\mu$ respectively.

The conventional atmospheric $\nu_e + \bar{\nu}_e$ flux is lower, implying that the prompt component is more visible in this channel.  We estimate that a measurement of the atmospheric $\nu_e + \bar{\nu}_e$ flux at $\sim$ 200 TeV at $\sim$ 50\% accuracy will cleanly distinguish between the pQCD contribution and intrinsic charm component.

The current upper limit on prompt neutrinos is 1.06 times the ERS w/G flux.  The neutrino flux due to intrinsic charm is at the same level as the ERS w/G flux implying that IceCube can constrain intrinsic charm of the proton.  This shows that IceCube can constrain QCD predictions in regions of parameter space which have been difficult to constrain in colliders for decades.

The multi-pronged approach consisting of IceCube data, collider physics, and global analysis will help us constrain the intrinsic charm of the proton, a $\sim$ 36 year old problem in QCD.  Using the weakly interacting neutrino to constrain the strong interactions also highlights the importance of cross disciplinary searches in physics.
 
\vspace{-0.5 cm}
	

\section*{Acknowledgments} 
We thank Markus Ahlers, Atri Bhattacharya, Alexander Friedland, Tom Gaisser, Joachim Kopp, Claudio Kopper, Shirley W. Li, Kenny C.Y. Ng, Mary Hall Reno, Juan Rojo, Carsten Rott, Jim Talbert, and Jakob van Santen for discussions.  We especially thank John F. Beacom, T.J. Hobbs, Matt Kistler, Ramona Vogt, and Tyce De Young for detailed discussions.  We thank the referees for a careful reading of the manuscript and for suggestions.  R.L. is supported by KIPAC.  S.J.B is supported by the Department of Energy, contract DE-AC02-76SF00515.  SLAC-PUB-16771	


\clearpage
\newpage

\begin{center}
\textbf{\large Supplemental Materials}
\end{center}

\begin{figure}
\centering
\includegraphics[angle=0.0,width=0.5\textwidth]{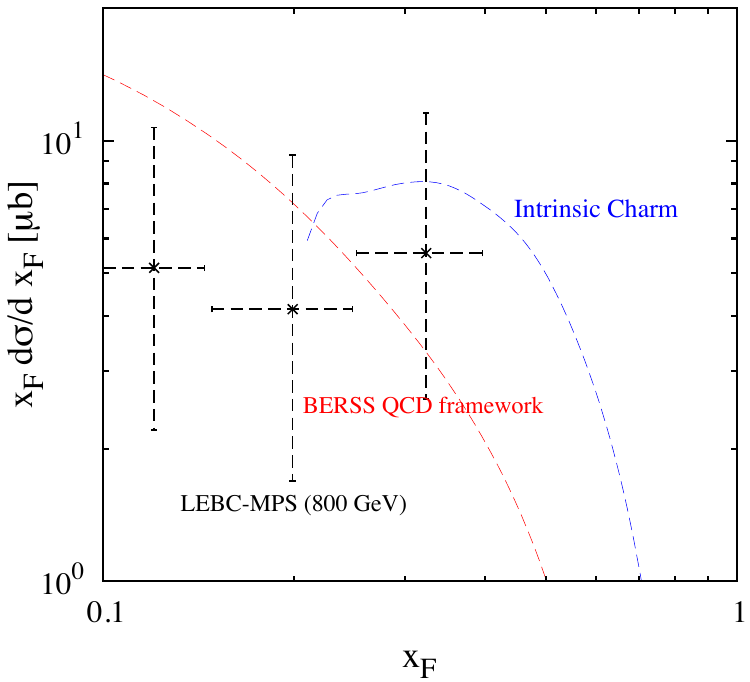}
\caption{The data at $x_F \gtrsim$ 0.1 as obtained by the LEBC-MPS collaboration\,\cite{Ammar:1988ta}.  We compare the data points with the best fit perturbative calculation estimated by us following Ref.\,\cite{Bhattacharya:2015jpa}.  We also show the intrinsic charm differential cross section for Case (A) and (B).  The differential cross section for Case (C) has the same shape, but a normalization which is $\sim$ 28\% of that shown in this figure.}   
\label{fig:LEBC-MPS}
\end{figure}

In this Supplementary Material, we show the $d\sigma/dx_F$ distribution that we use in our calculations.

The data points for $D/ \bar{D}$ production as measured by the LEBC-MPS collaboration\,\cite{Ammar:1988ta} for $x_F \gtrsim 0.1$ is shown in the Figure.  The beam energy in this fixed target experiment was 800 GeV.  We also show the best-fit pQCD framework prediction estimated by us following Ref.\,\cite{Bhattacharya:2015jpa}.  This best fit prediction is accompanied by a large error band because there is uncertainty in various parameters entering the pQCD calculations.  Given the large error bars in the measurement and the uncertainty in the theoretical prediction, it is difficult to reliably extract an intrinsic charm contribution from this measurement.  Our strategy is to use the best-fit prediction, and maximize the intrinsic charm contribution.  It is important to decrease the uncertainties in various parameters entering the pQCD calculation to reliably extract the intrinsic charm contribution\,\cite{Dulat:2013hea,Ball:2016neh}.

Although there are large error bars, the data points hint at a flat behavior with increasing $x_F$.  This is in contrast to the pQCD prediction which falls steeply.  Given this flattening hint and the large error bars, one can postulate an intrinsic charm contribution to the differential cross section.  The intrinsic charm differential cross section used in this work is such that it does not exceed the best - fit + 1-sigma measurement by the LEBC-MPS collaboration.  The contribution of intrinsic charm to the differential cross section is negligible compared to pQCD contribution for $x_F \lesssim$ 0.2 and we do not display the differential cross section in this region.  The intrinsic charm differential cross section as shown in the figure results in a total $d\sigma/dx_F \approx 35 \,\mu$b at $x_F \approx 0.32$.  This is the normalization of the intrinsic charm differential cross section that is used in Cases (A) and (B).

\begin{figure*}[!thpb]
\centering
\includegraphics[angle=0.0,width=0.49\textwidth]{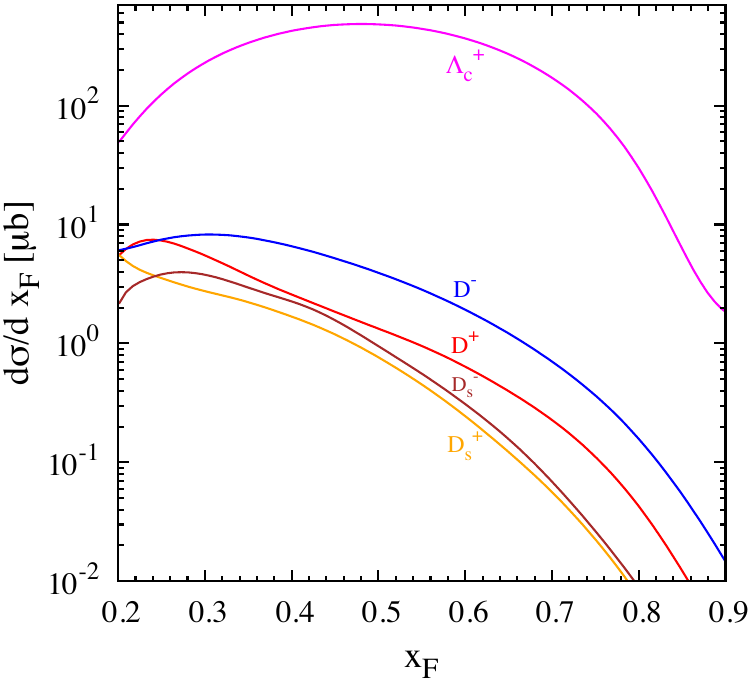}
\includegraphics[angle=0.0,width=0.49\textwidth]{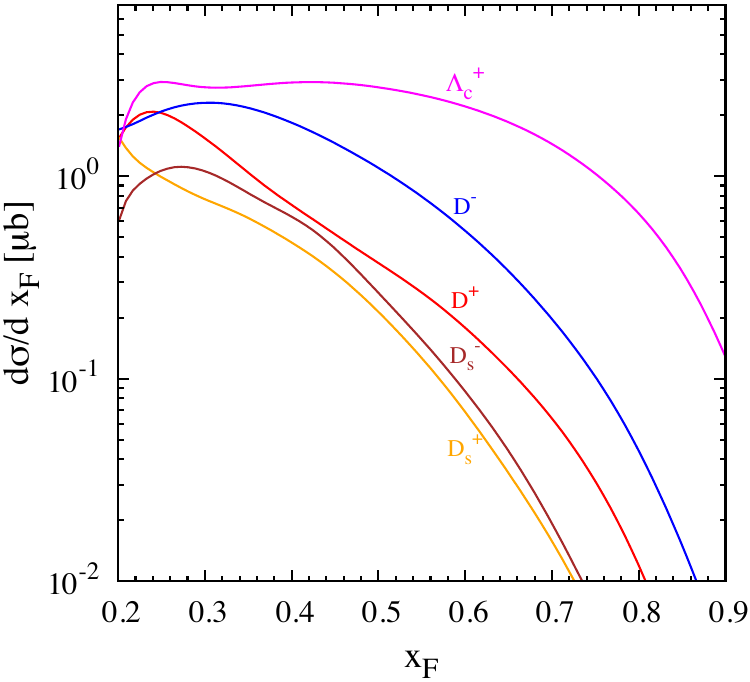}
\caption{The differential cross sections (at $\sqrt{s} \approx 39$ GeV) due to intrinsic charm employed in this work.  The left and right panel shows the differential cross sections employed in Case A and Case C respectively.  The differential cross sections employed in Case B is same as Case C but with an increased normalization by a factor of 3.57.  The differential cross section for $D^0$ and $\bar{D}^0$ are taken to be the same as $D^+$ and $D^-$ respectively.  Note that the scales of the y-axis in the two panels are different.}
\label{fig:Differential cross sections}
\end{figure*}

For Case (C) we use the same shape of the intrinsic charm differential cross section as before, but with a reduced normalization.  The normalization is this case is such that the total $d\sigma/d x_F \approx 25 \, \mu$b at $x_F \approx 0.32$.  The highest $x_F$ at which a measurement has been made by the LEBC-MPS collaboration is $x_F \approx 0.32$.  A future experiment which can measure the differential cross section at a higher $x_F$ will be invaluable for this field.

LEBC-MPS collaboration measure the cross sections for $D^\pm$, $D^0$, and $\bar{D}^0$ and present their differential cross sections for the sum of all these particles.  They do not discuss the measurement of $D_s^\pm$ in Ref.\,\cite{Ammar:1988ta}.  We estimate the contribution of $D_s^\pm$ by normalizing the contribution of $D^\pm$, $D^0$, and $\bar{D}^0$ from the model of Ref.\,\cite{Gutierrez:1998bc} to the data.  Since all the $D$ cross sections are related to each other, this gives us the contribution of $D_s^\pm$ which we use in all the three cases.

It is possible that intrinsic charm is smaller than what is assumed here.  In such a case, the intrinsic charm contribution to the atmospheric prompt neutrinos will decrease proportionately.  Due to the rigorous theoretical nature of intrinsic charm, some amount of intrinsic charm should be present in the proton.  Future experiments using different beam energies, and nuclear targets can measure the energy dependence and nuclear dependence of the intrinsic charm cross section more reliably.  Such a measurement will reduce most of the uncertainties in the calculations of the prompt neutrinos flux due to intrinsic charm.

The differential cross sections for the various charmed mesons and hadrons due to intrinsic charm for Case A and Case C are shown in Fig.\,\ref{fig:Differential cross sections}.


{\bf The case for Intrinsic Charm}  QCD predicts two distinct contributions to the heavy quark distributions in light hadrons such as the proton.  The  primary contribution comes from gluon splitting $g \to q \bar q$ -- the mechanism incorporated into the DGLAP pQCD evolution of structure functions. The resulting heavy-quark distribution falls as a power of $1-x$ faster than the gluon distribution,  and it is thus only important at low momentum fraction $x$.    

In addition to the gluon splitting mechanism, a second  contribution to the charm distribution $c(x,Q^2)$, which dominates at high $x$,  comes from QCD diagrams in which the heavy quark pair is attached by two or more gluons to the valence quarks of the proton;  it thus depends on the nonperturbative intrinsic structure of the proton\,\cite{Brodsky:1984nx,Brodsky:1980pb}.  Intrinsic charm has rigorous theoretical motivation and is a first principle prediction of QCD.  It can also be analyzed using OPE\,\cite{Brodsky:1980pb,Polyakov:1998rb}.  A typical intrinsic contribution comes from the $q \bar q$ cut of  the hadron self-energy diagrams analogous to light-by-light diagrams. One can use the operator product expansion to show that the resulting $q \bar q$  probability $P_{q \bar q}$ falls as $1/M^2_q$, in contrast to the $1/M^4_\ell$ behavior in Abelian QED\,\cite{Brodsky:1981se,Franz:2000ee}.  

The intrinsic charm distribution in the charm structure function of the proton is thus associated with the five-quark Fock state $|uud c \bar{c}\rangle$ of its light front wavefunction defined at fixed light-front time $\tau = t+z/c$ in the frame-independent eigensolution of  the QCD Light-Front Hamiltonian.   The probability distribution in invariant mass $dP_{q \bar q}/ M^2$  falls as $1/M^4$ and is thus  maximal in the proton light front wavefunction  at mininum off-shellness\,\cite{Brodsky:1984nx,Franz:2000ee}.   This occurs when all of the constituents in the hadron Fock state have the same rapidity; i.e.,  when they are all at rest in the parent hadron's rest frame.   Equal rapidity implies that the quark's light-front momentum $x = k^+/ P^+ $ is proportional to its transverse mass: $x_q = (m^2_q + k^2_\perp)/ (\sum^5_{i =1} m^2_q + k^2_\perp)$. Thus the heavy quarks  carry most of the LF momentum of the proton.  This key feature is incorporated by the BHPS model\,\cite{Brodsky:1980pb,Brodsky:1981se} for the intrinsic charm contribution to the $c(x,Q^2)$  and other heavy quark distributions. 

There are extensive indications for charm production  at high $x$, beginning with the EMC measurement of $c(x,Q^2)$ in deep inelastic muon scattering\,\cite{Aubert:1982tt}.   The rate observed by EMC is approximately 30 times higher at $x =0.42, Q^2$ = 75~GeV$^2$ than predicted by gluon splitting\,\cite{Harris:1995jx}.  Intrinsic charm also predicts the observed features of the data for ${d\sigma\over dx_F}(pp \to \Lambda_c X)$ as observed in ISR experiments\,\cite{Chauvat:1987kb} and more recently by SELEX\,\cite{Garcia:2001xj}.  In this case the comoving $c ,u$ and $d$ coalesce to produce the $ \Lambda_c$ at high $x_F$ where $x_F = x_c + x_u + x_d$.  Similarly, the $c$ and $\bar c$ can coalesce to produce the $J/\psi$ at high $x_F$\,\cite{Badier:1983dg}.  The production of two $J/\psi$ at high $x_F$  in the $\pi p \to J/\psi J/\psi X$  interaction as observed by NA3\,\cite{Badier:1982ae}, as well as the hadroproduction of double-charm baryons at high $x_F$, as observed by SELEX\,\cite{Ocherashvili:2004hi}  corresponds to  the materialization of the $|qqq c \bar c  c \bar c \rangle, $ and $|q \bar q c \bar c c \bar c \rangle$  Fock states of the incident hadrons\,\cite{Vogt:1995tf,Koshkarev:2016rci}.  Although there are large error bars, data from LEBC-MPS collaboration at 800 GeV at high $x_F$ ($x_F \gtrsim$ 0.1) on $D$/$\bar{D}$ (= $D^+ + D^0$ + antiparticles) production do not fall off as steeply as predicted in pQCD --- the flattening tendency hints at an intrinsic charm contribution.

Other high $x_F$ charm particle hadroproduction results are reviewed in Ref.\,\cite{Vogt:1994zf}.  None of these observations can be  explained by the ``color-drag"  model used in the PYTHIA simulations.  Even the modern version of PYTHIA does not include the effect of the intrinsic charm in the structure functions.  The corresponding $|uud b \bar{b} \rangle$ intrinsic bottom heavy-quark Fock state can account for the observation of $p p \to \Lambda_b X$ at high $x_F$ at  the ISR\,\cite{Bari:1991ty}.  The presence of charm at high $x$ in the proton structure function is also indicated by the anomalously large $p \bar p \to \gamma c X$ rate reported by the D0 experiment at the Tevatron\,\cite{Abazov:2009de,Mesropian:2014kfa}.  

An important test of $c(x,Q^2)$ at high $x$  can be performed at the LHC by measuring the production of the $Z^0$ boson at high $p_T$, balanced by a charm jet:  $p p \to Z^0 + c X$\,\cite{Boettcher:2015sqn,Lipatov:2016feu}.  Intrinsic heavy quark distributions also lead to  the production of the Higgs at high $x_F$ at the LHC\,\cite{Brodsky:2007yz}.  There are also proposals to perform a fixed target experiment at the LHC which will perfom high $x_F$ studies\,\cite{Brodsky:2012vg}.  There has been investigations about probing intrinsic charm via production of almost stationary doubly charmed baryons\,\cite{Groote:2017szb} and in LHCb and SMOG\,\cite{IltenAPS}.  Recent reviews and global analysis of intrinsic charm collider phenomenology are given in Refs.\,\cite{Brodsky:2015fna,Dulat:2013hea,Jimenez-Delgado:2014zga,Brodsky:2015uwa,Ball:2016neh}.  

Although there have been proposals to test intrinsic charm in LHC, it has not yet been demonstrated to work in practice.  More ways to test this important and uncertain component of QCD are needed.  We analyze the concept that IceCube can make substantial progress in increasing our knowledge about this component of QCD.

\bibliographystyle{kp}
\bibliography{Bibliography/references}	

\end{document}